\documentclass[11pt]{article}
\usepackage[dvips]{graphicx}
\usepackage{latexsym}
\usepackage[fleqn]{amsmath}

\newtheorem{theorem}{Theorem}

\newtheorem{proposition}{Proposition}
\newtheorem{corollary}{Corollary}

\def\squarebox#1{\hbox to #1{\hfill\vbox to #1{\vfill}}}
\def\qed{\hspace*{\fill}
        \vbox{\hrule\hbox{\vrule\squarebox{.667em}\vrule}\hrule}\smallskip}
\newenvironment{proof}{\begin{trivlist}
  \item[\hspace{\labelsep}{\em\noindent Proof.~}]
  }{\qed\end{trivlist}}

 \title{Quantum Evaluation of Multi-Valued Boolean Functions} 

   \author{Kazuo Iwama$^{\dag,  \dag\dag}$,  Akinori
     Kawachi$^{\dag,  \dag\dag}$,  
    Hiroyuki Masuda$^{\dag,  \dag\dag}$, \\
   Raymond H. Putra$^{\dag,  \dag\dag}$ and Shigeru Yamashita$^{\dag,
     \dag\dag\dag}$ \\    
   $^{\dag}$Quantum Computation and Information,  ERATO,  \\
   Japan Science and Technology Corporation (JST) \\
   $^{\dag\dag}$Graduate School of Informatics,  Kyoto University\\
   $^{\dag\dag\dag}$NTT Communication Science Laboratories\\ 
  E-mail: \{iwama,  kawachi,  hiroyuki, raymond, ger\}@kuis.kyoto-u.ac.jp.\\  
 }

\begin{document}







\setcounter{page}{1}
\date{}
\maketitle

%

\section*{Abstract}

Our problem is to evaluate a multi-valued Boolean function $F$ through oracle calls.
If $F$ is one-to-one and the size of its domain and range is the same,
then our problem can be formulated as follows: Given an oracle $f(a,x): 
\{0,1\}^n\times\{0,1\}^n \rightarrow \{0,1\}$ and a fixed (but hidden)
value $a_0$, we wish to obtain the value of $a_0$ by querying the oracle $f(a_0,x)$.
Our goal is to minimize the number of such oracle calls (the query complexity) using a quantum mechanism.

Two popular oracles are the EQ-oracle defined as $f(a,x)=1$ iff $x=a$ and the IP-oracle
defined as $f(a,x)= a\cdot x \mod 2$. It is also well-known that the query 
complexity is $\Theta(\sqrt{N})$ ($N=2^n$) for the EQ-oracle while only $O(1)$ 
for the IP-oracle. The main purpose of this paper is to fill this gap or to investigate
what causes this large difference. To do so, we introduce a parameter $K$ as the maximum
number of 1's in a single column of $T_f$ where $T_f$ is the $N\times N$ truth-table of the oracle $f(a,x)$.
Our main result shows that the (quantum) query complexity is heavily governed by this 
parameter $K$: ($i$) The query complexity is $\Omega(\sqrt{N/K})$. ($ii$) This lower bound 
is tight in the sense that we can construct an explicit oracle whose query complexity is 
$O(\sqrt{N/K})$. ($iii$) The tight complexity, $\Theta(\frac{N}{K}+\log{K})$, is also obtained 
for the classical case. Thus, the quantum algorithm needs a quadratically less number of
oracle calls when $K$ is small and this merit becomes larger when $K$
is large, e.g., $\log{K}$ v.s. constant when $K = cN$.


\section{Introduction}



\begin{figure}[tb]
\begin{center}
\resizebox{14cm}{!}{ 
\includegraphics*{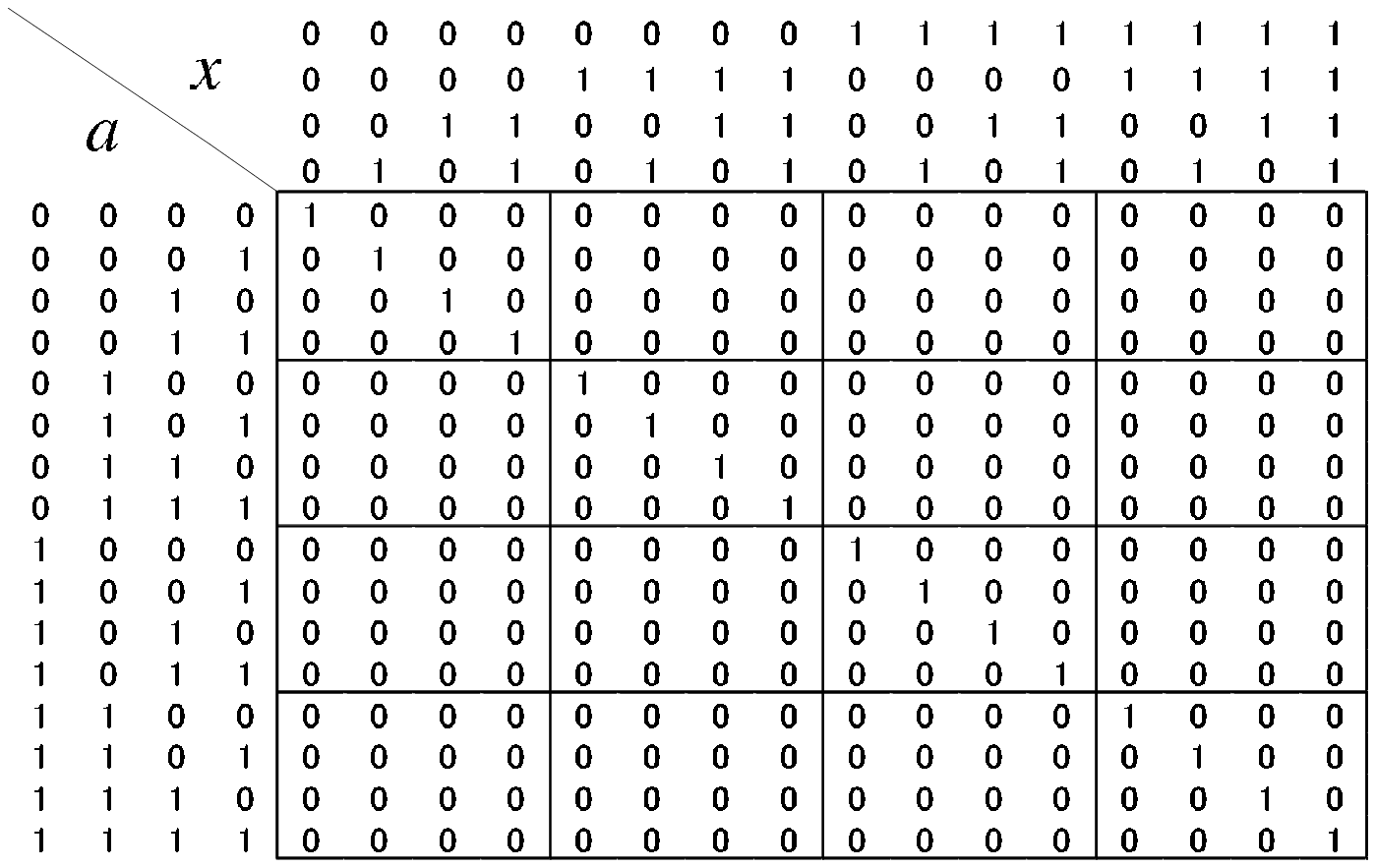} 
}
\caption{$f(a,x) = EQ_a(x)$}
\end{center}
\end{figure}

\begin{figure}[tb]
\begin{center}
\resizebox{14cm}{!}{ 
\includegraphics*{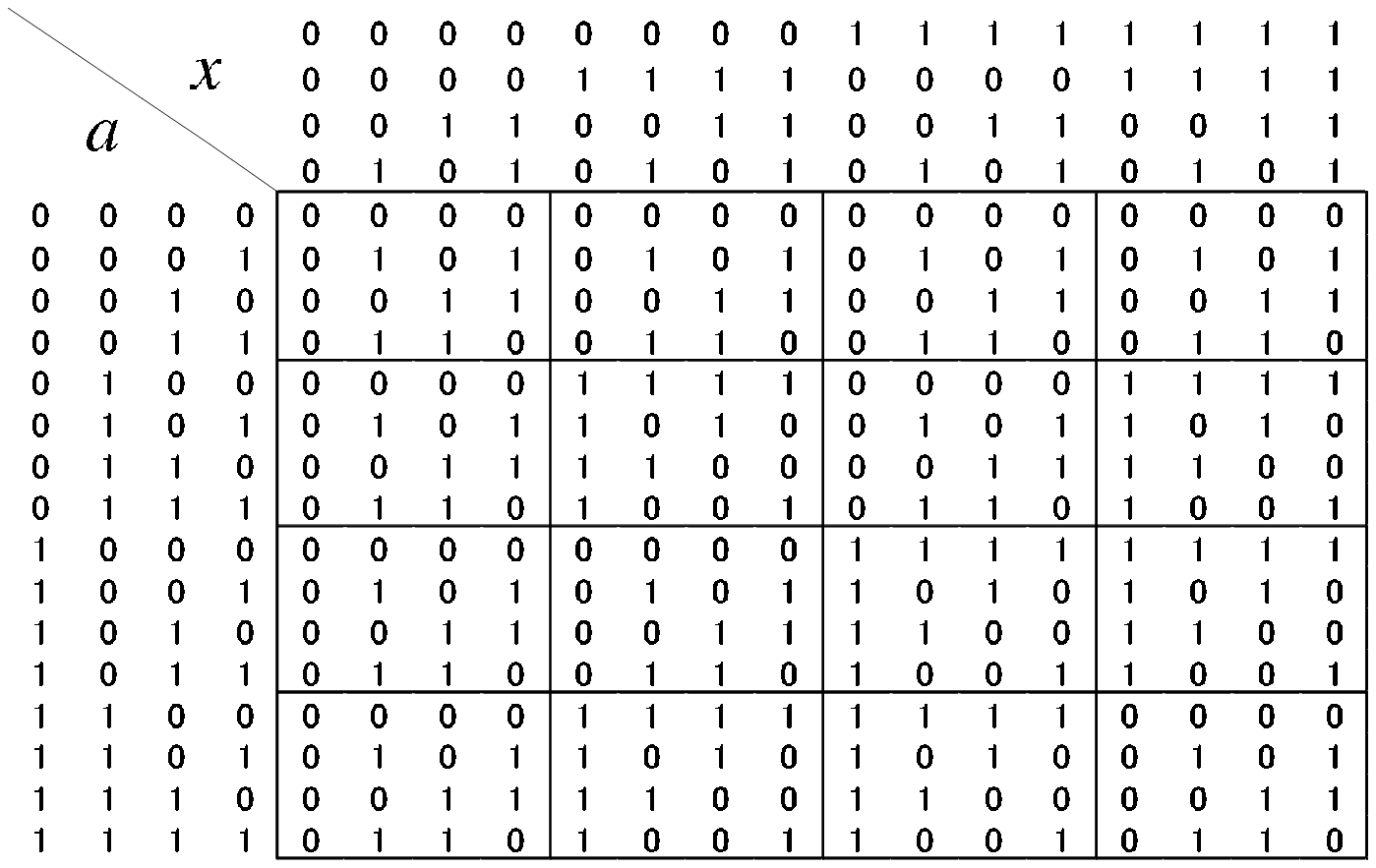} 
}
\caption{$f(a,x) = a\cdot x = \sum_i a_i \cdot x_i\quad\mod 2$}
\end{center}
\end{figure}

In \cite{Amb02}, Ambainis introduced the problem of evaluating a
(multi-valued and incompletely specified) Boolean function
$F(x_0,\ldots,x_{N-1})$ using a quantum mechanism. This problem includes
many interesting specific problems. For example, if we define 
$$
F(x_0,\ldots,x_{N-1}) = \begin{cases} a &\text{if $x_a = 1$ and $x_j =
0$ for all $j \neq a$},\\
\text{undefined} & \text{otherwise},\end{cases}
$$
then evaluating this function $F$ is the same as the so-called Grover
search. Also, if we define 
$$
F(x_0,\ldots,x_{N-1}) = \begin{cases} a &\text{if $x_i = a\cdot i$ mod 2 for
all $0 \le i \le N-1$},\\
\text{undefined} & \text{otherwise},\end{cases}
$$
then this is the same as the so-called Bernstein-Vazirani problem
\cite{Bernstein97} introduced in \cite{Gru99}. ``Quantum evaluation''
assumes that we can obtain the value of variable $x_i$ only through an oracle
$O(i)$. Since both functions are one-to-one, 
and their domain and range are of the same size, 
we can formulate the problem as follows. 

Let $n$ be an integer $\geq 1$ and $N = 2^n$. Then, given an oracle defined 
as a function 
$$
f(a,x):\{0,1\}^n \times \{0,1\}^n \rightarrow \{0,1\}
$$
such that $f(a_1,x) \neq f(a_2,x)$ for some $x$ if $a_1\neq a_2$,
and a fixed (and hidden) value $a$, we wish to obtain the value 
$a$, using the oracle $f(a,x)$. 
For the Grover search, one can easily see that the definition as 
$$
f(a,x) = \begin{cases} 1 &\text{if $x =a$},\\
0 &\text{otherwise},\end{cases}
$$
completely specifies the problem. This oracle is sometimes called the
EQ oracle and is denoted by $EQ_a(x)$. See Fig.~1 for $n =
4$. As illustrated in this figure, 
$f(a,x)$ is given by a truth table of size $N\times N$,
where each row gives the value of the function $F$ of the previous
definition. For example, we have $F(1,0,\ldots,0) = 0000$ from the
first row of the table. 
If the hidden value $a$ is $0010$ for example, the oracle returns value $1$ only when it 
is queried with $x=0010$. 
For the Bernstein-Vazirani problem, the similar definition is given as 
$$
f(a,x) = a \cdot x \quad\text{mod $2$},
$$
which is called the inner product (IP) oracle and denoted by
$IP_a(x)$. Its truth table for $n = 4$ is given in Fig.~2. 
(Through this paper, we assume that the domain 
of the Boolean function $F$ has the same size as its range. 
More general cases, e.g., the size of the range is larger than the domain,
will be mentioned briefly in the last section.)

As usual, our interest is in the ({\it quantum}) {\it query complexity}, i.e.,
how many oracle calls are needed to obtain the hidden value of
$a$. It is well-known that the query complexity for the EQ-oracle is
$\Theta(\sqrt{N})$\cite{Amb02, BBCMW98, BBBV97, BBHT96, Gro96, Vaz98}, while only $O(1)$ for the IP-oracle\cite{AC02, Bernstein97}. Why does such a
big difference exist between the EQ and IP oracles?
The difference might become clearer if we compare their truth tables
given in Figs. 1 and 2. One can immediately see that the table for
$IP_a$ is well-balanced in  terms of the numbers of $0$'s and $1$'s,
but quite unbalanced for $EQ_a$. The natural consequence is that there
should be intermediate oracles between those extreme ones for which
the query complexity is also intermediate between $\Theta(\sqrt{N})$
and $O(1)$. Furthermore these intermediate oracles could be
characterized by  some parameter in such a way that the query
complexity heavily depends upon this parameter value and both $EQ_a$
and $IP_a$ are obtained as special cases. The main purpose of this
paper is to show that this conjecture is true.  

{\bf Our Contribution.} Let $T_{f}$ be the truth-table of an oracle
$f(a,x)$ like the ones given in Figs.~1 and~2. We can assume 
without loss of generality that the number of $1$'s
is less than or equal to the number of $0$'s in each column of
$T_f$. Let $\#_i(T_{f})$ denote the number of $1$'s ($\le N/2$) in the $i$-th 
column of $T_{f}$ and $\#(T_{f}) = \displaystyle\max_i
\#_i(T_{f})$. Then we can 
show that this single parameter $\#(T_f)$ plays a key role, namely: ($i$) Let
$f(a,x)$ be any oracle and $K = \#(T_{f})$. Then the query 
complexity of the search problem for $f(a,x)$ is
$\Omega(\sqrt{N/K})$. For this result, we extend the main theorem in \cite{Amb02}.
Although the extension may seem moderate in terms of the statement, it actually 
increases a lot of usefulness of the theorem. 
($ii$) This lower bound is tight in the sense
that we can construct an explicit oracle whose query complexity is
$O(\sqrt{N/K})$. This oracle again includes both EQ and IP oracles
as special cases. ($iii$) The tight complexity,
$\Theta(\frac{N}{K}+\log{K})$, is also obtained for the classical 
case. Thus, the quantum algorithm needs a quadratically less number of
oracle calls when $K$ is small and the merit is even larger when $K$
is large, e.g., $\log{K}$ v.s. constant when $K = cN$.  

It should be noted that our (tight) bound $\Theta(\sqrt{N/K})$
looks similar to the tight bound of the $K$-solution Grover
search\cite{BBHT96}. This paper also examines the similarity and dissimilarity between
our problem and the multi-solution (MS) Grover: In our notation, 
MS-Grover is the problem for obtaining, for a fixed oracle $f(a,x)$ and
a fixed value $a$ both of which are hidden, the value $b \in \{0,1\}^n$ such that $f(a,b) =
1$. ($i$) There is an oracle for which our problem needs
$\Omega(\sqrt{N})$ oracle calls while MS-Grover $O(1)$ ones. ($ii$)
The complexity of the two problems is approximately the same if the
oracle is ``balanced'' (defined later) and $K \le N^{1/3}$. 
Intuitively our problem seems at least as hard as MS-Grover,
but we can also show the existence of an oracle which objects this intuition.  

{\bf Related Results.} Query complexity has constantly been one of the
central topics in quantum computation, especially in proving lower
bounds of quantum algorithms with oracles. Generally speaking, there are two
popular techniques to derive quantum lower bounds, i.e., polynomials and
adversary methods. The polynomials method was firstly introduced in
quantum computation by \cite{BBCMW98} who borrowed the idea from the
classical polynomial method. For example, it was shown that for
bounded error cases, evaluations of $AND$ and $OR$ functions need
$\Theta(\sqrt{N})$ number of queries, while parity and majority
functions at least $N/2$ and $\Theta(N)$, respectively. Recently, 
\cite{Aar02,Yi02} used the polynomials method to show the lower bounds
for collisions and element distinctness problems.

The adversary method was used in \cite{BBBV97,Vaz98}, although their
method resembles the classical one and for that reason it is called
the hybrid method. Their method can be used, for example, to show the
lower bound of the Grover Search.  Recently, Ambainis introduced a
quite general method, which is known as quantum adversary argument,
for obtaining lower bounds of various problems, e.g., the Grover
Search, AND of ORs and inverting a permutation\cite{Amb02}. Some lower
bounds are easier to obtain using the quantum adversary method than
the polynomials one. The most recent result is by \cite{BS02} who
extended \cite{Amb02} to establish a lower bound of $\Omega(\sqrt{N})$
on the bounded-error quantum query complexity of read-once Boolean
functions.

\section{Quantum Lower Bounds}


Our lower bound result is related to the main theorem of \cite{Amb02}. 
In this section, we give a slight
extension of the Ambainis' result, from which our result immediately
follows. We first review Theorem 5.1 in \cite{Amb02}. 
(Note that the statement uses the original definition of our problem.)

\begin{proposition}
Let $F(x_0, \ldots, x_{N-1})$ be a function of $N$ $\{0,1\}-$valued
variables and $X, Y$ be two sets of inputs such that $F(x) \neq F(y)$
if $x \in X$ and $y \in Y$. Let $R \subset X \times Y$ be such that \\
\hspace{6mm}1. For every $x \in X$, there exist at least $m$ different  
   $y \in Y$ such that $(x, y) \in R$. \\
\hspace{6mm}2. For every $y \in Y$, there exist at least $m'$ different  
   $x \in X$ such that $(x, y) \in R$. \\
\hspace{6mm}3. For every $x$ and $i \in \{0, \ldots, N-1\}$, there are at most $l$
different $y \in Y$ such that $(x, y) \in R$ and $x_i \neq y_i$. \\
\hspace{6mm}4. For every $y$ and $i \in \{0, \ldots, N-1\}$, there are at most $l'$
different $x \in X$ such that $(x, y) \in R$ and $x_i \neq y_i$. \\
Then, any quantum algorithm computing $F$ uses 
$\Omega(\sqrt{\frac{m m'}{l l'}})$ queries. 
\end{proposition}

Our extension is due to the fact that the selection of the two 
sets, $X$ and $Y$, may not necessarily be unique but may alter 
for each column. 

\begin{theorem}
Let $F(x_0, \ldots, x_{N-1})$ be a function of $N$ $\{0,1\}-$valued
variables and $X_i, Y_i$, for $i\in\{0, \ldots, N-1 \}$, be sets of inputs such
that $F(x) \neq F(y)$ 
if $x \in X_i$ and $y \in Y_i$, and $X_0 \cup Y_0=\cdots=X_{N-1} \cup Y_{N-1}$. Let $R_i \subset X_i \times Y_i$ be
such that  \\
\hspace{6mm}1. For every $x \in X_i$, there exist at least $m$ different  
   $y \in Y_i$ such that $(x, y) \in R_i$. \\
\hspace{6mm}2. For every $y \in Y_i$, there exist at least $m'$ different  
   $x \in X_i$ such that $(x, y) \in R_i$. \\
\hspace{6mm}3. For every $x$ and $j \in \{0, \ldots, N-1\}$, there are at most $l$
different $y \in Y_i$ such that $(x, y) \in R_0 \cup \cdots \cup R_{N-1}$ and $x_j \neq y_j$. \\
\hspace{6mm}4. For every $y$ and $j \in \{0, \ldots, N-1\}$, there are at most $l'$
different $x \in X_i$ such that $(x, y) \in R_0 \cup \cdots \cup R_{N-1}$ and $x_j \neq y_j$. \\
Then, any quantum algorithm computing $F$ uses 
$\Omega(\sqrt{\frac{m m'}{l l'}})$ queries. 
\end{theorem}

\begin{proof}
Omitted since it is enough to make word-to-word replaces in the
original proof of \cite{Amb02}.
\end{proof}

Now our lower-bound result is almost immediate. Recall that our oracle
is defined by 
 $f(a,x): \{0,1\}^n \times \{0,1\}^n \rightarrow \{0,1\}$. If we use
 the notation $F$, its range is $\{0,1\}^n$ and for any $x, y \in
\{0,1\}^N$ such that $x \neq y$, it is guaranteed that $F(x) \neq
F(y)$ if they are defined.  

\begin{corollary}
Let $f$ be an oracle such that $\#(T_{f}) = K$.  
Then any quantum algorithm that computes a target $a$ with probability 
$1-\varepsilon$ needs  $\Omega(\sqrt{N/K})$ oracle calls for
some fixed $\varepsilon < 1/2$.   
\end{corollary}

\begin{proof}  
Recall that $K \leq N/2$. So we can select 
 $X_i$ and $Y_i$ of Theorem~1 such that $|X_i| = |Y_i| = N/2$ and if
 $x_i = 1$ then $x \in X_i$ (i.e., all $Y_i$ do not include $y$ such
that $y_i=1$).  
Now it is not hard to see that we can set $m=m'= l = N/2$ and $l'=K$. 
Hence, Theorem~1 implies that the query complexity is 
$\Omega(\sqrt{\frac{m m'}{l l'}}) = 
\Omega(\sqrt{\frac{(N/2)^2}{NK/ 2}})=
\Omega(\sqrt{\frac{N}{K}})$.
\end{proof}

Note that Proposition 1 is, if we use it as it is, not so powerful;
we cannot prove the lower bound of the Grover search for instance.
In \cite{Amb02}, Ambainis made a different kind of extension against 
Proposition 1 to prove several lower bounds including Grover's.
This extension is quite powerful but does not seem appropriate for 
our purpose. Our extension is also powerful, which will be used later 
to prove other lower bounds.


\section{Quantum Upper Bounds}
In this section, we show that there exists a specific oracle whose query 
complexity is $O(\sqrt{N/K})$. 
The oracle intuitively works partly as the EQ-oracle and partly as the
IP-oracle and therefore we call it a {\it hybrid oracle}. 
The hybrid oracle with a parameter $1\leq k \leq n$, denoted by $h_k(a,x)$, is
defined as follows ($k$ indicates the size of its IP-part): Let
$a=(a_1, a_2,\ldots, a_{n-k}, a_{n-k+1},\ldots, a_{n})$
and $x=(x_1, x_2,\ldots, x_{n-k}, x_{n-k+1},\ldots, x_{n}).$ Then
$h_k(a,x)=1$ iff 
(i) $(a_1,..., a_{n-k})=(x_1,..., x_{n-k})$ and 
(ii) $(a_{n-k+1},...,a_{n})\cdot (x_{n-k+1},..., x_{n}) =0$ (mod $2$). Fig.~3
shows $h_2$ for $n=4$. One can see that the small $4\times 4$ matrix is the negation of
the IP-oracle.
\begin{theorem}
There exists a quantum algorithm to find a target $a$ using $O(\sqrt{2^{n-k}})$ calls of $h_k$.
\end{theorem}

\begin{figure}[tb]
\begin{center}
\resizebox{14cm}{!}{ 
\includegraphics*{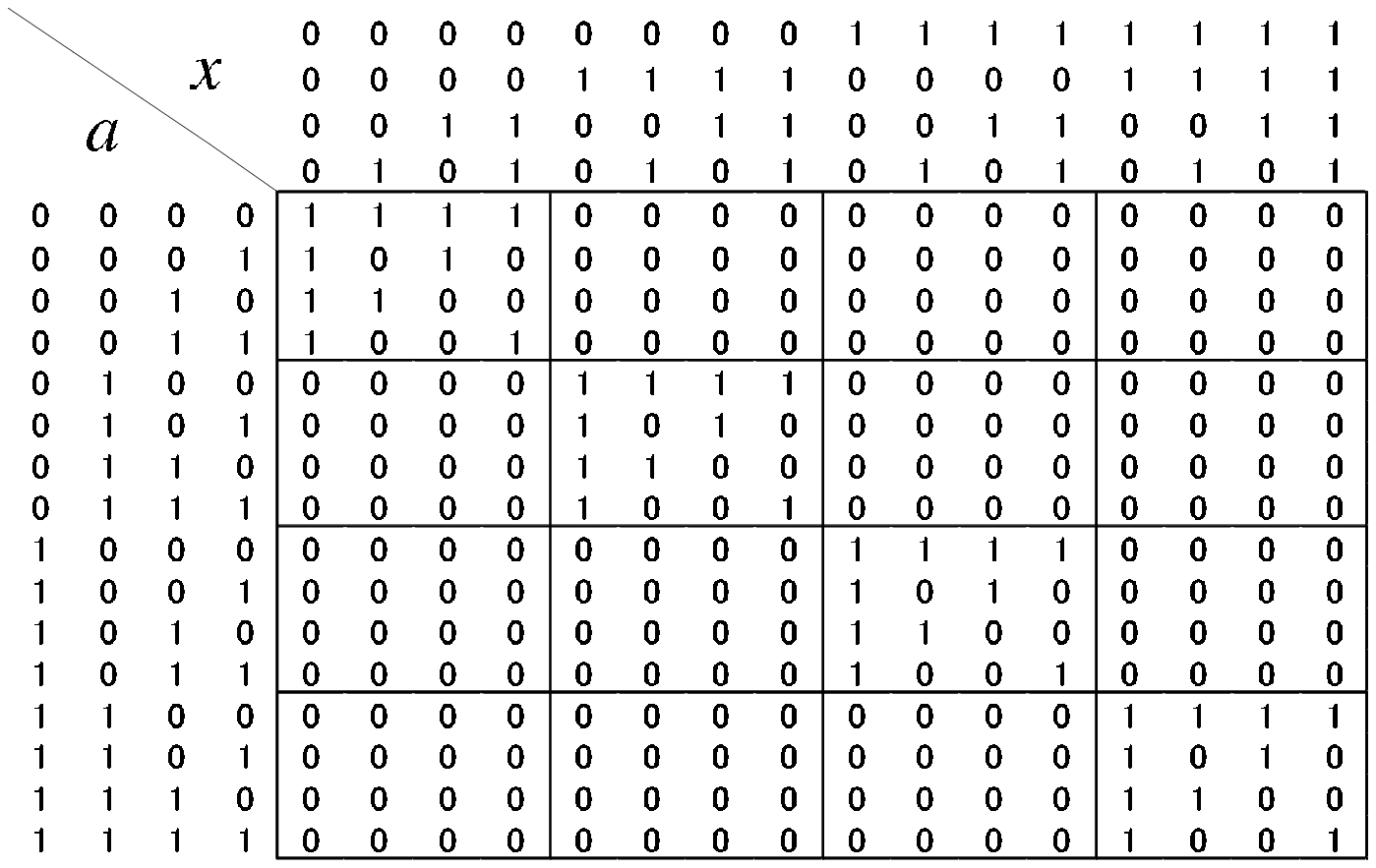} 
}
\caption{The hybrid oracle with $n=4$ and $k=2$}
\end{center}
\end{figure}

\begin{proof} 
We combine Grover search\cite{Gro96, BBHT96} with BV
algorithm\cite{Bernstein97} to find a target $a$ using the hybrid oracle
$h_k$. At first, we can determine the first $n-k$ bits of $a$. Fixing the
last $k$ bits to $|0 \rangle$, we apply Grover search using oracle $h_k$
for the first $n-k$ bits to determine $a_1,...,a_{n-k}$.
 It should be noted that $h_k(a, (a_1, \ldots, a_{n-k}, 0, \ldots, 0))
= 1$ and $h_k(a, (x_1, \ldots, x_{n-k}, 0, \ldots, 0))=0$ for any
$x_1,...,x_k \neq a_1,...,a_k$.
Next, we apply BV algorithm to determine the remaining $k$ bits.
This algorithm requires $O(\sqrt{N/K})$ queries for Grover search and 
$O(1)$ queries for BV algorithm. 
\end{proof}

\begin{corollary}
There exists an oracle $f$ such that $\#(T_f)=K$ and its query complexity is $O(\sqrt{N/K})$.
\end{corollary}


Thus Corollary 1 is tight in the sense that we cannot improve it in general.
Note that both Corollaries 1 and 2 hold for the EQ- and IP-oracles.
We can also prove the same upper bound, $O(\sqrt{N/K})$, for a more general 
class of oracles if we restrict the value of $K$. An oracle $f$ is said to be
{\it balanced} if all columns and all rows have the same number of 1's.

\begin{theorem} 
For a balanced oracle $f(a,x)$ with $K \leq N^{1/3}$, there exists a
quantum algorithm to find a target $a$ using $O(\sqrt{N/K})$ queries
of $f$. 
\end{theorem}
\begin{proof} 
Since $f(a,x)$ is balanced, each row and each column of $T_f$ has 
exactly $K$\ 1's. 
Our algorithm consists of two parts; the first part is quantum and 
the second part classical:  
The first part of the algorithm is to reduce the size of the search
space from $N$ to $K$ which is achieved by using (multi-solution) Grover search. It first
finds an $x$ such that $f(a,x) = 1$. Since 
each row of $T_f$ has $K$\ 1's, the query complexity of this first part is
$O(\sqrt{N/K})$. 
Also since each column has $K$\ 1's, the number of $a$ satisfying $f(a,x) = 1$ for a
particular $x$ is exactly $K$. Now, let a subset $R = \{a|f(a,x) = 1\}$, 
which includes the candidates of $a$. 

The second part of the algorithm is to repeatedly choose an index $y$
such that $\exists a,b \in R$ where $f(a,y) \neq f(b,y)$. Clearly,
unless $|R| = 1$, there must exist such $y$, 
which can reduce the number of the candidates at least by one. 
More formally, let the query's result for such 
$y$ is $m \in \{0,1\}$. Then, we can substitute $R \cap
\{b|f(b,y) = m\}$ for $R$, by which $|R|$ decreases at least by one.
The total query complexity is thus $O(\sqrt{N/K}) + K$, which is $O(\sqrt{N/K})$ 
for $K \leq N^{1/3}$. 
\end{proof}


\section{Relation to Grover Upper and Lower Bounds}
As described in Section~1, our problem is closely related to MS-Grover.
However, there exist some gaps between those two problems. 
In this section, we show some explicit oracles such that 
the query complexities of those two problems are quite different.
The following theorem shows the oracle such that our problem is harder than MS-Grover.
\begin{theorem} There is an oracle $f$ such that the query complexity of our search problem is $\Omega(\sqrt{N})$ and 
that of MS-Grover is $O(1)$.
\end{theorem}
\begin{proof} Consider an oracle $f$ whose truth-table is such that the
upper left and the lower right of the table is exactly the same as the
EQ-oracle, and all the elements of the upper right and the lower left
of the table are 1's. (See Fig.~4 for $N=16$.) Then it is obvious 
that we need only $O(1)$ queries to find $x$ such that $f(a,x)=1$
since the number of such $x$ is so large that even choosing $x$ randomly suffices. 
On the other hand, we can show the hardness of finding $a$ for this
oracle by Theorem 1. We select $X_i$ and $Y_i$ of Theorem 1 such that $X_i \cup Y_i$ is
equal to the upper half of the truth-table, $|X_i|=|Y_i|=N/4$ and if
$x_i=1$ then $x \in X_i$ for $i \leq N/2$.
Then, one can easily see that $m=m'=l=N/4$ and $l'=1$. Therefore, we 
need $\Omega(\sqrt{N})$ queries to find $a$. 
\end{proof}

\begin{figure}[tb]
\begin{center}
\resizebox{14cm}{!}{ 
\includegraphics*{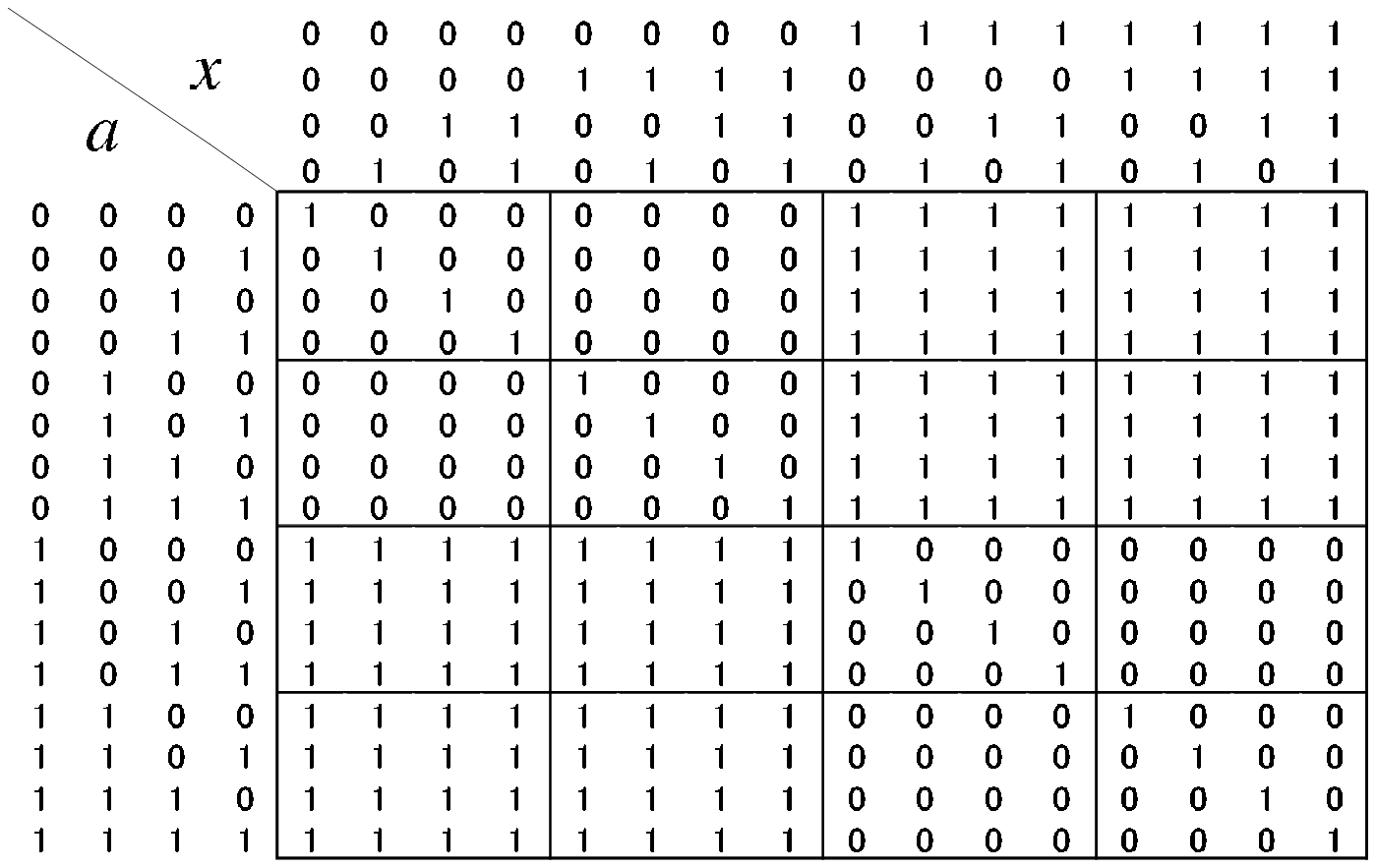} 
}
\caption{A truth-table where our problem is harder than MS-Grover}
\end{center}
\end{figure} 

Thus our problem is much harder than MS-Grover in this particular oracle. 
It is also true, as shown in the previous section, that our problem has 
approximately the same query complexity as MS-Grover for balanced oracles with
$K \leq N^{1/3}$. 
Then, is there any oracle for which MS-Grover is {\it harder} than our problem? 
We have no formal answer to this question, but the following example suggests 
that the answer is probably yes: 
Let $IP_{a}'$ be an oracle which is exactly the same as $IP_a$ except 
that $f(0,z)=1$ for some single $z$ (recall that $f(0,z)=0$ in $IP_a$).
Then, if we simply use the Grover search, it needs $\sqrt{N}$ queries when $a=0$. 
(Recall that we assumed that MS-Grover does not have any information 
on the oracle table $T_f$ so that this seems to be the best we can do.) 
In contrast, if we use the BV algorithm, then we can get 
the value of $a$ with high probability since $IP_{a}'$ is so close to $IP_{a}$. 
(It is easy to compute the success probability, which may be omitted.)


\section{Classical Lower and Upper Bounds}

We first give the classical lower bound and then prove the matching upper bound using the same oracle given in the previous section.

\begin{theorem}
Suppose that $f$ is an arbitrary oracle such that $\#(T_f)$ is $K$. Then any classical algorithm needs at least 
$\lfloor\frac{N}{K}\rfloor+\lfloor\log{K}\rfloor-2$ oracle calls to obtain a target in the worst case.
\end{theorem}

\begin{proof}
The following proof is due to the standard adversary argument. Let $A$ be any classical algorithm using the oracle $f$. 
Suppose that the target is $a \in \{0,1\}^n$. Then the execution of $A$ is described as follows:
(i) In the first round, $A$ calls the oracle with the predetermined value $x_0$ and the oracle answers with $d_0=f(a,x_0)$. 
(ii) In the second round, $A$ calls the oracle with value $x_1$, which is determined by $d_0$ and the oracle answers with $d_1=f(a,x_1)$. 
(iii) In the $(i+1)$-st round, $A$ calls the oracle with $x_{i}$ which is determined by $d_0,d_1,...,d_{i-1}$ and 
the oracle answers with $d_{i} = f(a, x_{i})$.
(iv) In the $m$-th round $A$ outputs the target $a$ which is determined by
$d_0, d_1,..., d_{m-1}$ and stops. Thus, the execution of $A$ is
completely determined by the sequence $(d_0, d_1,..., d_{m-1})$
which is denoted by $A(a)$. (Obviously, if we fix a specific target
$a$, then $A(a)$ is uniquely determined).

Let $m_0 = \lfloor N/K \rfloor + \lfloor \log K \rfloor -3$ and suppose that $A$
halts in the $m_0$-th round. We compute the sequence
$(c_0,c_1,\ldots,c_{m_0}),\, c_i\in\{0,1\}$, and another sequence 
$(L_0,L_1,\ldots,L_{m_0}),\, L_i\subseteq \{a|a\in\{0,1\}^n\}$, as follows (note
that $c_0,\ldots,c_{m_0}$ are similar to $d_0,...,d_{m-1}$ above and are chosen by the adversary): (i) $L_0 = \{0,1\}^n$.
(ii) Suppose that we have already computed $L_0,...,L_i$, and $c_0,...,c_{i-1}$. Let $x_i$ be the value
with which $A$ calls the oracle in the $(i+1)$-st round.
(Recall that $x_i$ is determined by $c_0,...,c_{i-1}$.)
Let $L^0 = \{s\, |\, f(s,x_i) =0\}$ and $L^1 = \{s\, |\, f(s,x_i)=1\}$.
Then if $|L_i \cap L^0| \geq |L_i \cap L^1|$ then we set $c_i = 0$ and $L_{i+1} = L_i \cap L^0$.
Otherwise, i.e., if $|L_i\cap L^0|<|L_i\cap L^1|$, then we set $c_i = 1$ and $L_{i+1} = L_{i} \cap L^1$.

Now we can make the following two claims.

\textbf{Claim 1.} $|L_{m_0}|\geq 2$. 
(Reason: Note that $|L_0|=N$ and the size of $L_i$ decreases as $i$
increases. By the construction of $L_i$, one can see that until
$|L_i|$ becomes $2K$, its size decreases additively by at most $K$ in
a single round and after that it decreases multiplically at most one
half. The claim then follows by a simple calculation.) 

\textbf{Claim 2.} If $a \in L_{m_0}$, then $(c_0,\ldots ,c_{m_0})=A(a)$.
(Reason: Obvious since $a\in L_0\cap L_1\cap \cdots\cap L_{m_0}$.)

Now it follows that there are two different $a_1$ and $a_2$ in $L_{m_0}$ such that $A(a_1)=A(a_2)$
by Claims 1 and 2. Therefore $A$ outputs the same answer for two
different targets $a_1$ and $a_2$, a contradiction. 
\end{proof}


Next, we consider classical upper bounds for the hybrid oracle
discussed in Section 3. Recall that $K = 2^k$. 

\begin{theorem}
There exists a classical algorithm to find a target $a$ using
$O(\frac{N}{K}+\log{K})$ calls of $h_k$. 
\end{theorem}

\begin{proof} The algorithm consists of an exhaustive and a binary search.
First, we determine the first $n-k$ bits by fixing the last $k$ bits to all $0$'s and using exhaustive search. 
Second, we determine the last $k$ bits by using binary search.
This algorithm needs $2^{n-k} (=\frac{N}{K})$ queries in the exhaustive search, 
and $O(k)(=O(\log{K}))$ queries in the binary search. 
Therefore, the total complexity of this algorithm is $O(2^{n-k}+ k)
(=O(\frac{N}{K}+\log{K}))$. 
\end{proof}
\section{Concluding Remarks}
We have assumed through this paper that the size of $T_f$ is $N \times N$, i.e., 
the size of the domain and the size of the range of the Boolean function $F$ are the same. 
This restriction can of course be removed. 
For example, consider the 
$
\left(
\begin{array}{c}
N \\
K
\end{array}
\right) 
\times N
$ 
table $T_f$ which includes every string $y$ consisting of $K~1$'s and 
$(N-K)~0$'s in its rows. 
One can see this is exactly the same as the $K$-solution Grover search if 
our answer is to obtain an $x$ such that $f(a,x)=1$ and its complexity 
is $\Theta(\sqrt{\frac{N}{K}})$. 
Recall again that our present problem is different, 
i.e., we need to obtain $a$. 
We can also use Theorem~1, which implies (details are omitted) 
a lower bound of $\Omega(\sqrt{\frac{N}{K}})$. 
This lower bound is the same as the lower bound of MS-Grover and seems 
too weak as a lower bound of obtaining $a$. 
Fortunately it is not hard to improve this lower bound up to 
$\Omega(\sqrt{N})$ if $K=cN$: 
By selecting $K$ rows appropriately, we can obtain a smaller table 
which is similar to the table given in Fig.~4. 
Nevertheless, this $\sqrt{N}$ lower bound still seems weak; 
the real bound is probably much larger than $\sqrt{N}$. 
To obtain such a ``{\it huge}" lower bound for natural oracles 
(\cite{Amb02} has obtained one for an artificial oracle) 
should be interesting future work.

Another future direction is to find 
a different parameter which can control {\it upper} bounds of the query
complexity of evaluating functions. Note that the Hamming distance
between two rows of $T_f$ is $O(1)$ for the EQ-oracle and $N/2$ for
the IP-oracle. This fact might be a good hint for this direction. 

\end{document}